\documentclass[10pt,conference]{IEEEtran}

\usepackage[nointlimits]{amsmath}
\usepackage{amsthm}
\usepackage{amssymb, cancel}
\usepackage{graphicx, nicefrac}
\usepackage{color}
\usepackage[dvipsnames]{xcolor}
\usepackage{txfonts}
\usepackage{mathtools}
\usepackage{hyperref}
\usepackage{multirow}
\usepackage{tikz}
\usepackage{float}
\hypersetup{colorlinks=true, citecolor=blue, linkcolor=blue}
\graphicspath{{./figures/}}
\theoremstyle{definition}
\newtheorem{definition}{Definition}

\usepackage{url}

\def\ket#1{{\left| #1 \right\rangle}}

\begin{document}

\title{Predictive Models from Quantum Computer Benchmarks}
\author{\IEEEauthorblockN{Daniel Hothem\IEEEauthorrefmark{1}\IEEEauthorrefmark{4}, Jordan Hines\IEEEauthorrefmark{1}\IEEEauthorrefmark{2} Karthik Nataraj\IEEEauthorrefmark{3}, Robin Blume-Kohout\IEEEauthorrefmark{1} and Timothy Proctor\IEEEauthorrefmark{1}}

\IEEEauthorblockA{\IEEEauthorrefmark{1}Quantum Performance Laboratory,
Sandia National Laboratories, Livermore, CA 94550, USA}
\IEEEauthorblockA{\IEEEauthorrefmark{2}Department of 
Physics, University of California, Berkeley, CA 94720}
\IEEEauthorblockA{\IEEEauthorrefmark{3}Institute for Computational and Mathematical Engineering, Stanford University, Stanford, CA 94305}
\IEEEauthorblockA{\IEEEauthorrefmark{4}Email: dhothem@sandia.gov}
}
\date{\today}
\maketitle

\begin{abstract}
    Holistic benchmarks for quantum computers are essential for testing and summarizing the performance of quantum hardware. However, holistic benchmarks---such as algorithmic or randomized benchmarks---typically do not predict a processor's performance on circuits outside the benchmark's necessarily very limited set of test circuits. In this paper, we introduce a general framework for building predictive models from benchmarking data using \emph{capability models}. Capability models can be fit to many kinds of benchmarking data and used for a variety of predictive tasks. We demonstrate this flexibility with two case studies. In the first case study, we predict circuit (i) process fidelities and (ii) success probabilities by fitting \emph{error rates models} to two kinds of volumetric benchmarking data. Error rates models are simple, yet versatile capability models which assign effective error rates to individual gates, or more general circuit components. In the second case study, we construct a capability model for predicting circuit success probabilities by applying transfer learning to ResNet50, a neural network trained for image classification. Our case studies use data from cloud-accessible quantum computers and simulations of noisy quantum computers.
\end{abstract}

\section{Introduction}\label{sec:introduction}

Quantum processors have rapidly grown over the past decade, but hardware errors (i.e., noise) limit their computational capabilities. The errors in one- or two-qubit systems can be studied in detail using tomographic methods \cite{Nie21}, but contemporary processors suffer from complex errors (e.g., crosstalk) that are challenging to fully characterize \cite{sarovar2019detecting}. This has led to the proliferation of holistic benchmarks that aim to quantify the overall impact of errors on a processor's performance \cite{Cro19, Blu20, Pro21, Lub23}. Holistic benchmarks run suites of test circuits and use the data to compute metrics that summarize a processor's performance, e.g., mean gate infidelities \cite{Mag11, Pro22, Hin22}, the quantum volume \cite{Cro19}, volumetric benchmarking plots \cite{Blu20}, or capability regions \cite{Pro21}. However, while many holistic benchmarks offer useful summaries of a processor's performance, most benchmarks do not make \emph{predictions} about how well other circuits or algorithms will execute. Any performance predictions from the results of a holistic benchmark have been typically obtained using informal and \emph{ad hoc} extrapolations (e.g., see \cite{Lub23}). 

In this work we introduce a framework for constructing predictive models from benchmarking data, and we provide two case studies demonstrating how to do so. Our general framework (see Section~\ref{sec:our-method}) is based on \textit{capability models}, which builds on the concept of a processor's capability function \cite{Hot23}. A capability function captures how well a processor can run circuits---by formalizing the concept of a circuit's ``success probability'' using a performance metric such as process fidelity---and a capability model is simply a parameterized model for a capability function. As we explain herein, most holistic benchmarks can be interpreted as probing a capability function, and so the parameters of a capability model can be fit to benchmarking data. Capability models are flexible, and we provide two case studies to demonstrate the promise of this approach. 
 
Our first case study (see Section~\ref{sec:erm-case-study}) introduces \emph{error rates models} (ERMs), which are flexible, scalable, and interpretable capability models that can be designed to predict a variety of figures of merit. ERMs generalize and formalize the widely-used idea of representing the errors in each of a processor's gates with a generic error process (global depolarization) that has a single parameter---the gate's error rate. Fitting ERMs to benchmarking data  summarizes the data in terms of \emph{effective} error rates and enables predictions for how other circuits will perform. Our second case study (see Section~\ref{sec:resnet50-case-study}) creates a capability model for circuit success probabilities from a pre-trained neural network. We apply transfer learning \cite{Wei16} to ResNet50 \cite{He15}, an image classifier, to create a capability model that predicts circuit performance on a simulated noisy quantum computer. This complements recent work that has used custom (rather than pre-trained) neural networks to predict a variety of circuit performance metrics \cite{Hot23, Ame22, Liu20, Vad22, Wan22}.

\section{Predictive models from benchmarks} \label{sec:our-method}

\subsection{Benchmarks and capability functions}\label{sec:background}
We begin with a brief overview of holistic quantum computer benchmarks, and we explain how most benchmarking data can be interpreted as estimates of a \emph{capability function} \cite{Hot23}. Holistic benchmarks typically consist of: (i) selecting some set of circuits $\{c\}$, e.g., via sampling from some distribution; (ii) running these circuits (or some closely related circuits) on the processor being tested; (iii) for each circuit $c \in \{c\}$, computing from the data a single number $\hat{s}(c)$ that is an estimate of a metric $s(c)$ (such as fidelity) that quantifies how well the processor can run $c$; and (iv) computing one or more summary statistics from the data $\{\hat{s}(c)\}_{c \in \{c\}}$. We refer to $s(c)$ as the benchmark's \emph{capability function} \cite{Hot23}. 

Examples of benchmarks that can be described as above include many RB methods \cite{Mag11, Pro22, Hin22}, the quantum volume benchmark \cite{Cro19}, cross-entropy benchmarking \cite{Box19}, volumetric benchmarks \cite{Blu20}, and many algorithmic benchmarks \cite{Lub23}. For example, many RB methods consist of running randomly sampled circuits where each circuit's overall action is to map the standard input state $\ket{00\cdots}$ to some computational basis state $\ket{x}$. Therefore, an RB circuit is a \emph{definite outcome} circuit, meaning that it produces one particular bit string $x$ if it is run without error. In RB, each circuit's success probability---i.e., the probability the correct bit string is returned---is estimated from data. So this benchmark's capability function $s(c)$ is success probability.

\subsection{Capability models}
\begin{figure}
    \centering    \includegraphics[width=5cm]{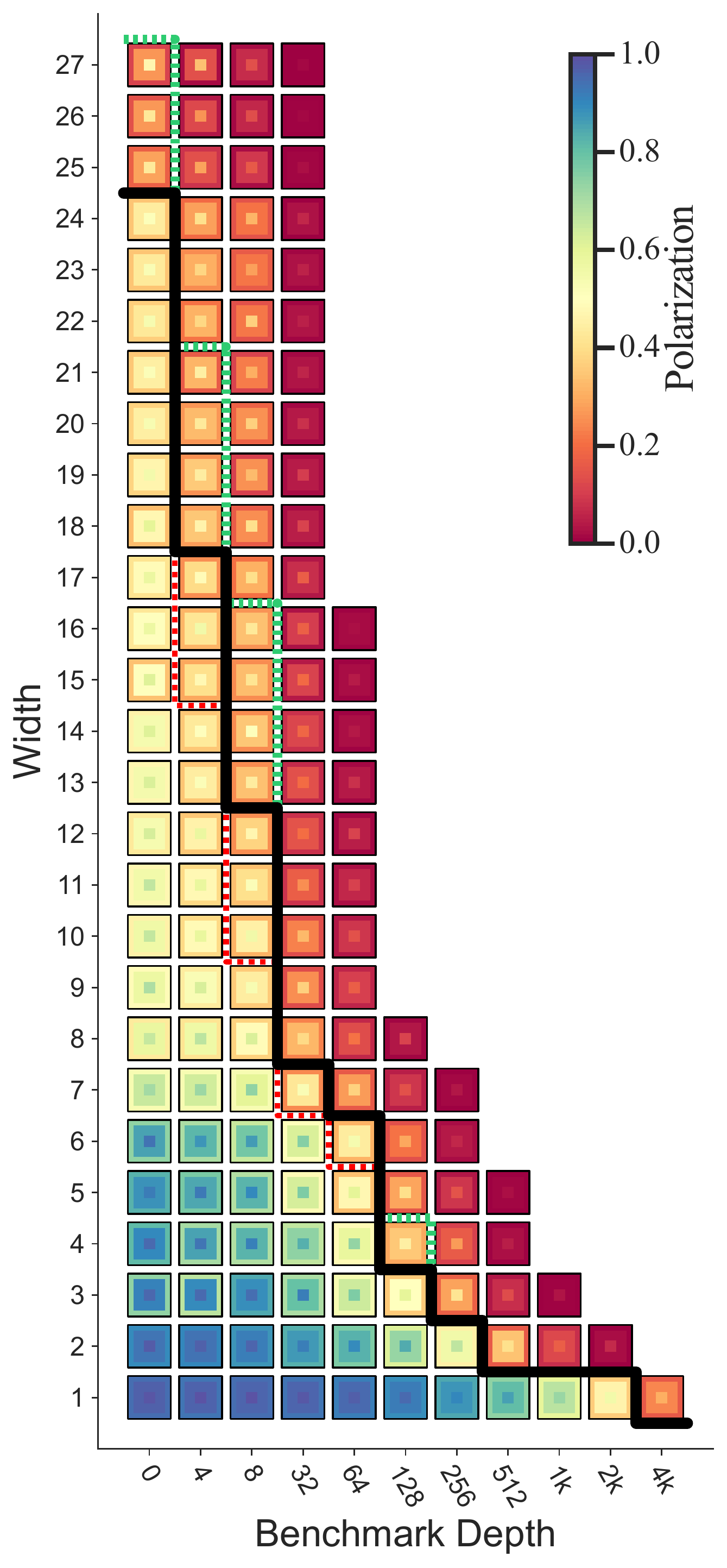}
    \caption{\small{\textbf{Volumetric benchmarks are not predictive}. A volumetric benchmarking plot \cite{Blu20} summarizing data from randomized mirror circuits (RMCs) run on \texttt{ibmq\_montreal}. For each circuit width and benchmark depth (see definition in Ref.~\cite{Pro21}), the concentric squares show the maximum (inner square), mean (middle square), and minimum (outer square) of the estimated polarizations for all the circuits of that shape that were run. The polarization ($s_{\textrm{pol}}$) of the output distribution of a definite outcome $n$-qubit circuit $c$ is simply $s_{\textrm{pol}} = (s_{\textrm{sp}} -\nicefrac{1}{2^n})/(1-\nicefrac{1}{2^n})$ where $s_{\textrm{sp}}$ is $c$'s success probability \cite{Pro21}. Frontiers (green, black, and red lines) show the circuit shapes at which these three statistics drop below the threshold value of $\nicefrac{1}{e}$. Volumetric plots summarize a processor's observed performance on some (necessarily small) set of benchmarking circuits $\{c\}$, but they make no predictions. In this work, we propose fitting parameterized models to benchmarking data like this, enabling predictions of the processor's performance on other circuits.}}
    \label{fig:ibm_montreal_vb}
\end{figure}

Holistic benchmarks summarize their data using one or more statistics or plots---e.g., see the volumetric benchmarking plot of Fig.~\ref{fig:ibm_montreal_vb}--- but they typically do not make predictions. We propose fitting parameterized models to benchmarking data to construct predictive models from benchmarks. Our framework is based on a particular kind of parameterized model that we call a \emph{capability model}.
\begin{definition}\label{def:error-model}
    Let $s$ be a capability function $s: \mathcal{C} \rightarrow \mathbb{R}$ defined over a set of circuits $\mathcal{C}$.
    A capability model for $s$ is a parameterized function $\epsilon: \mathcal{C} \rightarrow \mathbb{R}$ used to approximate $s$.
\end{definition}
Capability models do \emph{not} predict the outcome distribution of a circuit, unlike many error models for quantum processors (such as those constructed via gate set tomography \cite{Nie21}). Instead capability models predict only how accurately the processor implements circuits, as quantified by a capability function $s$.  A capability model can be any type of parameterized function (e.g., a neural network) and its parameters do not need to correspond to the rates of physical errors (e.g., gate over-rotation angles etc.).

We propose building predictive models from benchmarks as follows:
\begin{enumerate}
    \item[(1)] Run a benchmark that generates data $\{\hat{s}(c)\}$ for a set of circuits $\{c\}$, where $\hat{s}(c)$ is an estimate of some capability function $s(c)$.
    \item[(2)] Select a capability model $\epsilon$ for capability function $s$.
    \item[(3)] Fit the parameters of $\epsilon$ to the data $\{\hat{s}(c)\}$.
\end{enumerate}

We will explore two kinds of capability models in this work: ERMs (see Section~\ref{sec:erm-case-study}) and pre-trained neural networks (see Section~\ref{sec:resnet50-case-study}). 

\section{Case study 1: Error rate models} \label{sec:erm-case-study}
In this case study, we introduce error rates models (ERMs) and demonstrate their utility by fitting them to benchmarking data.

\subsection{Theory}\label{ssec:erm-case-study:theory}
ERMs generalize and formalize two simple concepts: (i) represent the error in each of a processor's quantum logic operations by a single error rate, and (ii) predict a circuit's failure (or success) rate by a simple function (e.g., the sum) of the error rates of all the operations in the circuit.
\begin{definition}\label{def:erm}
    An error rates model $E$, for a capability function $s:\mathcal{C}\to\mathbb{R}$, is a capability model defined by the tuple $(\mathcal{X},\mathcal{E}, \mathcal{N}, f)$ where:
    \begin{itemize}
        \item  $\mathcal{X}$ is a set of quantum logic operations, called basis elements;
        \item $\mathcal{E} = \lbrace \epsilon_{x} \in \mathbb{R} \mid x \in \mathcal{X}\rbrace$ are the model's parameters, called error rates;
        \item  $\mathcal{N}: \mathcal{C}\to \mathbb{N}^{|\mathcal{X}|}$ counts how many times each basis element $x \in \mathcal{X}$ appears in a circuit, defined by a rule $\mathcal{R}$ for decomposing any circuit $c \in \mathcal{C}$ into the basis elements $\mathcal{X}$;
        \item and $f: \mathbb{N}^{|\mathcal{X}|} \rightarrow \mathbb{R}$ is an $\mathcal{E}$-parameterized function that is composed with $\mathcal{N}$ to compute the model's prediction, i.e., $E(c) = f[\mathcal{N}(c)]$.
    \end{itemize}
\end{definition}

ERMs are simple and flexible. For example, an ERM's basis elements can contain a diverse set of circuit components---such as gates, circuit layers, or even large subroutines used in algorithms. ERMs are also scalable, i.e., it is feasible to fit ERMs to many-qubit data.

We now introduce a class of ERMs that are based on \emph{global depolarization}, which is a simple and widely-used model for errors in quantum gates \cite{Pro21, Hin22, Hot23, qiskit}. Consider a circuit set $\mathcal{C}$ and corresponding basis element set $\mathcal{X}$ that contains (i) circuit sub-components (e.g., one- and two-qubit gates) that ideally implement unitary evolutions, and (ii) readout operations that only occur at the end of a circuit (so, e.g., $\mathcal{X}$ does not include mid-circuit measurements). Now model the imperfect operation of each $x \in \mathcal{X}$ in an $n$-qubit circuit by the perfect operation preceded by an $n$-qubit depolarizing channel\footnote{For example, if $\mathcal{X}$ contains one- and two-qubit gates, the error channel for a $3$-qubit circuit layer containing a one-qubit gate ($x_1$) and a two-qubit gate ($x_2$) is the product of two $3$-qubit depolarizing channels with infidelities $\epsilon_{x_1}$ and $\epsilon_{x_2}$. Because $n$-qubit depolarizing channels commute the order in which the error channels are applied is irrelevant.} ($  \mathcal{D}_{\epsilon_x}$) with process infidelity $\epsilon_x$, i.e.,
\begin{equation}
    \mathcal{D}_{\epsilon_x}[\rho] = \gamma_x \rho + [1-\gamma_x] \mathbb{I}/2^n,
\end{equation}
where $ \gamma_x=\gamma(F_x)$, $F_x = 1-\epsilon_x$ is the \emph{process fidelity} with which operation $x$ is implemented, and
\begin{equation}
    \gamma(F) = \frac{4^nF - 1}{4^n-1}, \label{eq:pol}
\end{equation}
is a rescaling of process fidelity called \emph{process polarization}. Because $n$-qubit depolarizing channels commute with each other and every $n$-qubit unitary superoperator, this model predicts that the superoperator implemented when executing a circuit $c$ is simply
\begin{equation}
    \Lambda(c) = \left(\prod_{x \in \mathcal{R}(c)} \mathcal{D}_{\epsilon_x}\right) \mathcal{U}(c) =  \left(\prod_{x \in \mathcal{X}} \mathcal{D}_{\epsilon_x}^{\mathcal{N}(c)}\right) \mathcal{U}(c) , \label{eq:lambda-dep}
\end{equation}
where $\mathcal{U}(c)$ is the superoperator representation of the unitary ideally implemented by $c$, and $\mathcal{N}_x(c)$ is the number of instances of basis element $x$ in $c$. Given a capability function $s$ (e.g., process fidelity) and a basis element set $\mathcal{X}$, this model for $c$'s superoperator implies an ERM for $s$. We now demonstrate ERMs based on Eq.~\eqref{eq:lambda-dep}.

\subsection{Example 1: Predicting process fidelities}\label{ssec:erm-case-study:demonstration-1}
Most benchmarks run a set of circuits $\{c\}$ and then evaluate performance on each circuit $c$ by the difference between the observed and ideal output distributions of $c$ (quantified using, e.g., cross-entropy \cite{Box19}, classical fidelity \cite{Lub23}, or heavy output probability \cite{Cro19}). However, stricter measures of performance can be obtained by quantifying the difference between the actual and ideal quantum processes implemented when running $c$ \cite{Proctor2022-zs}, using, e.g., process fidelity ($F$)---or, equivalently, process polarization [Eq.~\eqref{eq:pol}]. Benchmarks whose data $\{\hat{s}(c)\}$ consists of estimates of process polarization (or fidelity)
for a circuit set $\{c\}$ can be constructed using mirror circuit fidelity estimation (MCFE) \cite{Proctor2022-zs}. MCFE is an efficient method for estimating the process fidelity of a circuit $c$ by embedding $c$ within a variety of other circuits.

\begin{figure}
    \centering \includegraphics[width=8cm]{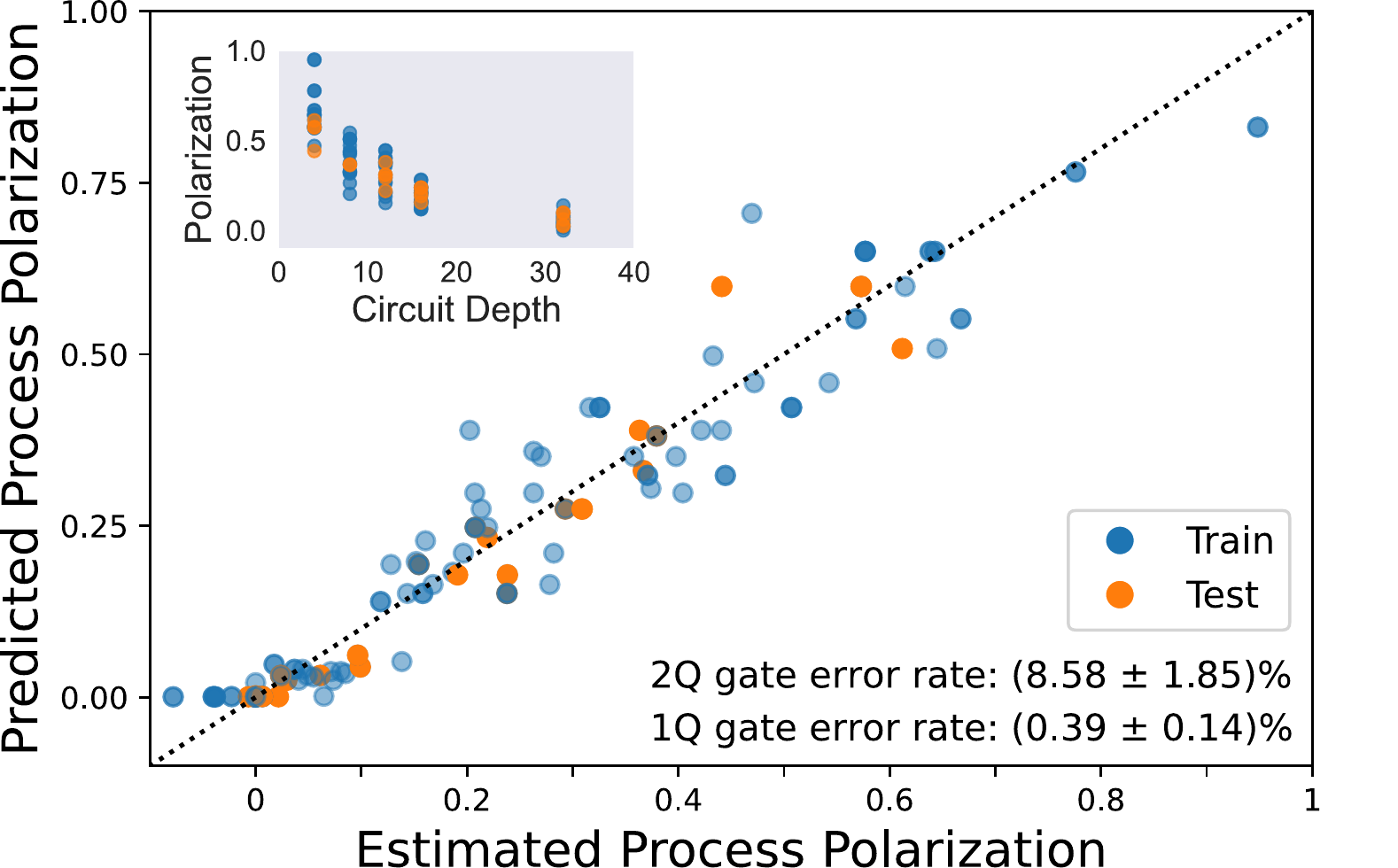}
    \caption{\small{\textbf{Predicting circuit process fidelities}. The predictions of a two-parameter ERM for \texttt{ibmq\_kolkata} that was fit to benchmarking data consisting of estimated process polarizations (a rescaling of process fidelity) for 150 random circuits. The ERM was fit to training data (120 circuits, blue points) and assessed on holdout test data (30 circuits, orange points). The data is summarized in the inset, which shows estimated process polarization versus circuit depth. The fit values for the model's parameters (lower right)---an error rate for one-qubit gates and an error rate for two-qubit gates---are effective error rates that summarize the data.}}
    \label{fig:kolkata-polarization}
\end{figure}

We now demonstrating fitting ERMs based on the global depolarizing model [Eq.~\eqref{eq:lambda-dep}] to benchmarking data that has process polarization as its capability function. The model of Eq.~\eqref{eq:lambda-dep} implies that the process polarization of a circuit $c$ is
\begin{equation}
    E(c) = \prod_{x \in \mathcal{X}} \gamma_x^{\mathcal{N}_x(c)}.
\end{equation}
We fit this capability model to data from \texttt{ibmq\_kolkata} that consists of estimated process polarizations for 150 12-qubit random circuits (these estimates were obtained using MCFE). The inset of Fig.~\ref{fig:kolkata-polarization} summarizes the data. We fit a simple ERM, with two basis elements---a one-qubit gate and a two-qubit gate---and corresponding error rates. 

Figure~\ref{fig:kolkata-polarization} shows the predictions of the best-fit ERM (fit with a least squares objective function). We find effective one- and two-qubit error rates of $\epsilon_1 = (0.39\pm 0.14)\%$ and $\epsilon_2 = (8.5 \pm 1.8)\%$, respectively. These best-fit error rates are approximately eight times larger than the mean of IBM's reported one- and two-qubit gate error rates for the twelve qubits used in this experiment ($\epsilon_1 = 0.05\%$ and $\epsilon_2 = 1.1\%$) and our best-fit error rates are a better description of the data: the mean absolute error $\delta_{\textrm{abs}}(E) = \mathbb{E}_c|E(c)-\hat{s}(c)|$ of our two-parameter ERM with best-fit error rates is $3.0\%$ whereas if we use IBM's error rates it is $11.8\%$.

\begin{figure}[!h]
    \centering    
    \includegraphics[width=8.5cm]{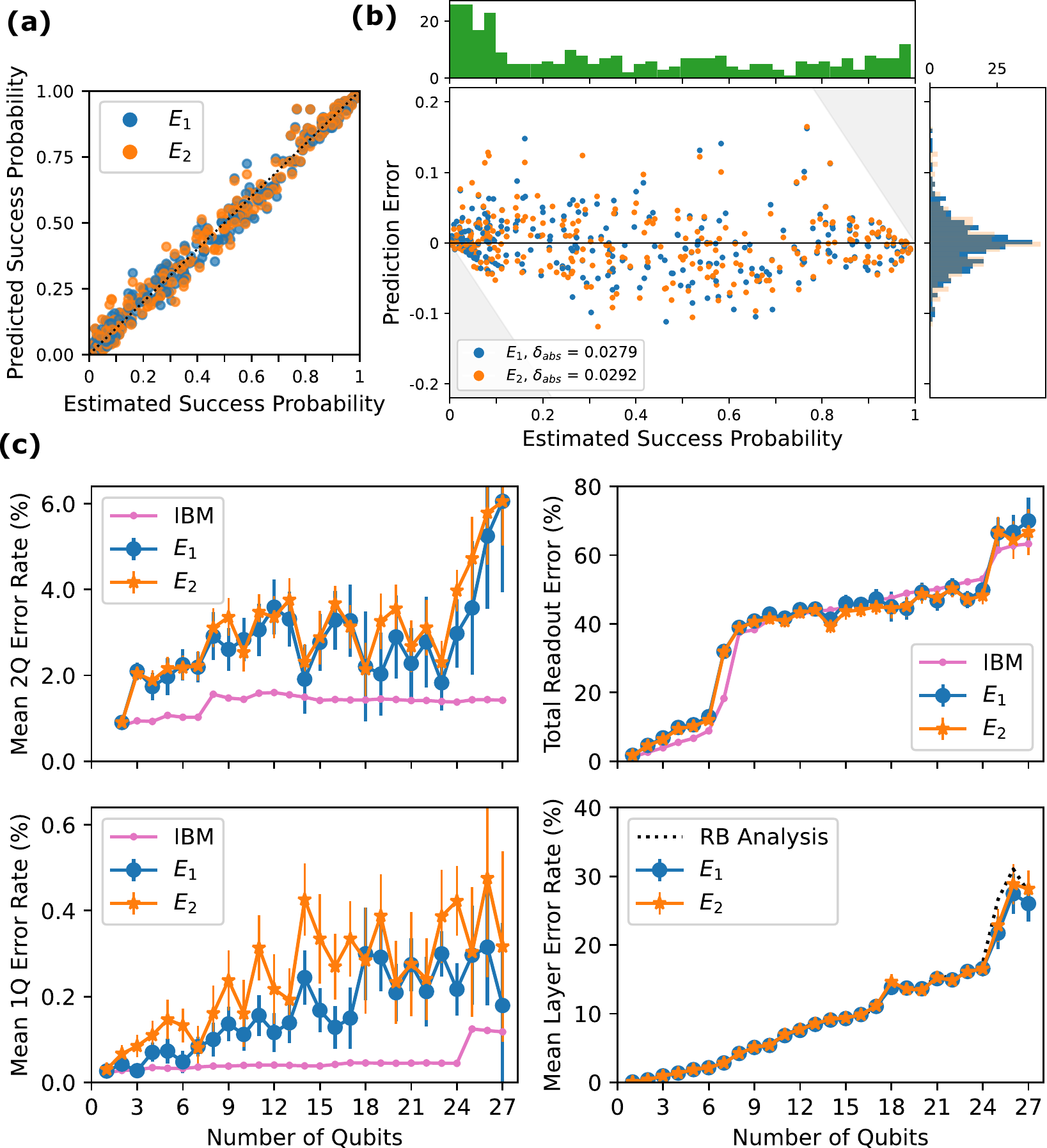}
    \caption{\small{\textbf{Effective error rates for \texttt{ibmq\_montreal}}. We fit ERMs to 1-27 qubit random circuit data from \texttt{ibmq\_montreal}. \textbf{(a)} The predictions for holdout test data of two ERMs $E_1$ (blue) and $E_2$ (orange) (details in the main text) versus each circuit $c$'s success probability $\hat{s}(c)$. Both models have moderate prediction accuracy, as summarized by \textbf{(b)} prediction errors, i.e., $\delta(c) = E(c) - \hat{s}(c)$ (see legend for the mean absolute error). \textbf{(c)} Both ERMs are parameterized by error rates that depend on circuit width. Here we show each model's average two-qubit gate error rate (top left), average one-qubit gate error rate (lower left) and total readout error (upper right) versus the circuit width (number of qubits) alongside IBM's reported error rates. IBM's average gate error rates fail to account for increasing crosstalk errors as circuit width increases, and our gate error rates are \emph{effective error rates} that better describe the data. Each ERM's error rates can be used to compute a mean error rate ($\bar{\epsilon}_w$) for a random $w$-qubit circuit layer (lower right). We observe close agreement between each ERM's estimate for $\bar{\epsilon}_w$ and an independent estimate extracted from the data by a conventional RB analysis (at each $w$, we fit the success probabilities versus circuit depth to an exponential). All uncertainties are $1\sigma$ and are computed using a bootstrap.}}
    \label{fig:erm-predictions}
\end{figure}

\subsection{Example 2: Predicting circuit success probabilities}\label{ssec:erm-case-study:demonstration-2}
Many benchmarks (e.g., RB) run definite outcome circuits and estimate these circuits' success probabilities---the probability that the circuit returns its single correct bit string. The capability model for the success probability of an $n$-qubit definite outcome circuit $c$ implied by Eq.~\eqref{eq:lambda-dep} is simply
\begin{equation}
  E(c) =  \left(1 - \frac{1}{2^n}\right)\prod_{x \in \mathcal{X}} \gamma_x^{\mathcal{R}(c)_x} + \frac{1}{2^n}.
\end{equation}

We demonstrate this family of capability models using data from randomized mirror circuits (RMCs) \cite{Pro22} run on \texttt{ibmq\_montreal}, a 27-qubit system. We ran approximately 3000 circuits with varied circuit widths and depth. For each RMC we estimate its success probability.\footnote{RMCs data is often analyzed using so-called \emph{adjusted} success probabilities defined in Refs.~\cite{Pro22, Hin22}, but we do not this here for conceptual simplicity.} The data is summarized in Fig.~\ref{fig:ibm_montreal_vb}. For each width $w$, circuits were run on a single connected set of $w$ qubits ($\Gamma^w$) and these sets were nested ($\Gamma^w \subset \Gamma^{w+1}$). We fit two ERMs ($E_1$ and $E_2$) with different basis elements $\mathcal{X}$. To investigate crosstalk errors in \texttt{ibmq\_montreal} using ERMs we fit models in which each gate's error rate is indexed by circuit width. We describe this in terms of sub-ERMs: Each ERM $E_i$ consists of 27 different sub-ERMs ($E_i^w$) where $E_i^w$ makes predictions for (and is fit to the data from) circuits of width $w$. Each sub-ERM in $E_1$ is a three-parameter model, with generic one-qubit gate, two-qubit gate, and readout basis elements. In contrast, each sub-ERM in $E_2$ models one-qubit gates (two-qubit gates) on different qubits (qubit pairs) with independent error rates.

Figure~\ref{fig:erm-predictions} shows (a) the predictions and (b) prediction errors, of best-fit $E_1$ and $E_2$ capability models. These models were fit to training data (90\% of the circuits) using maximum likelihood estimation. In Figure~\ref{fig:erm-predictions} (a-b) we evaluate the models using holdout test data (the remaining 10\% of the circuits). Both models have moderately low prediction error---the mean absolute errors on the test data are $\delta_{\textrm{abs}}(E_1) \approx 2.8\%$ and $\delta_{\textrm{abs}}(E_2) \approx 2.9\%$. The additional parameters of $E_2$ therefore do not improve prediction accuracy on the test data. The fit error rates---the parameters of $E_1$ and $E_2$---are summarized in Fig.~\ref{fig:erm-predictions}~(c). The average one-qubit gate and two-qubit gate error rates of $E_1$ and $E_2$ (which depend on circuit width $w$) are in close agreement. Both one- and two-qubit gate error rates generally increase with $w$, becoming significantly larger than the average gate error rates reported by IBM [compare the blue and orange points with the pink points in the left column of Fig.~\ref{fig:erm-predictions}~(c)]. This discrepancy suggests large crosstalk errors, as IBM's gate error rates do not include the effects of crosstalk (they are estimated using one- and two-qubit RB). Our gate error rates are \emph{effective error rates} that better describe the data and include the impact of crosstalk. In contrast, our total readout error estimates are consistent with IBM's reported readout error rates.

\section{Case study 2: Transfer learning with Resnet50}\label{sec:resnet50-case-study}
In our second case study, we create a capability model by applying transfer learning to ResNet50, a pre-trained neural network.

\subsection{Neural network capability models}
Neural networks are highly-expressive parameterized models that are general-purpose function approximators~\cite{Hor89}.
Recent work has explored using custom neural networks as capability models with promising results~\cite{Hot23, Ame22, Vad22, Wan22}. However, training and tuning \emph{de novo} neural network capability models is costly---it can be computationally intensive and can require large amounts of data. One approach to reducing the cost of creating neural network capability models is \emph{transfer learning}. Transfer learning is a broad set of techniques designed to modify pre-trained neural networks for use on a new task~\cite{Wei16}. Transfer learning can be particularly valuable when training data for the new task is scarce.

\subsection{Capability models from ResNet50}
To explore the feasibility of creating capability models from pre-trained neural networks, we used transfer learning techniques to create a capability model from ResNet50 \cite{He15}. ResNet50 is a large pre-trained image classifier that consists of 48 convolutional layers, an average pooling layer, a max pooling layer, and a 1000-unit classification layer~\cite{He15}. We modified ResNet50 to create a capability model by replacing ResNet50's final 1000-unit classification layer with a single-unit dense layer that has a sigmoid activation function (so its predictions are within $[0,1]$). Only the weights in the final single-unit dense layer are learnable parameters, i.e., all the weights in the layers from ResNet50 are fixed (frozen).

\subsection{Encoding circuits for ResNet50}
Inputting a circuit $c$ into ResNet50 requires a representation $I(c)$ of $c$ that is compatible with ResNet50. As an image classifier, ResNet50 processes three-dimensional tensors. We therefore input circuits into ResNet50 by modifying the three-dimensional tensor encoding $I'(c)$ of circuit introduced in Ref.~\cite{Hot23}. This encoding represents a $w\times d$ circuit $c$ for an $n$-qubit device as an $n \times d_{\textrm{max}}$ color image where $d_{\textrm{max}}$ is the depth of the deepest circuit in the dataset, i.e., $I'(c)$ is an $n\times d_{\textrm{max}} \times h$ tensor where $h$ is the number of ``color'' channels ($h = 10$ for the encoding of Ref.~\cite{Hot23}). The color channels store information about which gate is performed on each qubit in each layer of the circuit (as well as some limited information about each qubit's error sensitivity). See Ref.~\cite{Hot23} for details on the circuit encoding $I'(c)$. 

ResNet50 accepts tensors with three channels, so we must map $I'(c)$ to a three-channel image $I(c)$. Our mapping consists of simply reshaping $I'(c)$ (after embedding it within a larger tensor, if necessary). This reshaping destroys some locality information encoded within $I'(c)$, and more principled or trainable $I'(c)\to I(c)$ mappings are possible.

\subsection{Demonstration on simulated data}\label{ssec:resnet50-case-study:data}
To train and test our ResNet50-based capability model we used simulated data from a publicly-accessible repository~\cite{dataset-hot23}. The data consists of success probabilities for RMCs for a 5-qubit processor. The circuits ranged in width from 1 to 5 qubits and in depth from 3 to 1825 layers. The circuits were simulated with a stochastic Pauli errors model (detailed in Section IV A of Ref.~\cite{Hot23}). We used training, validation, and test datasets containing 996, 664, and 1494 circuits, respectively.

\begin{figure}
    \centering    \includegraphics[width=8cm]{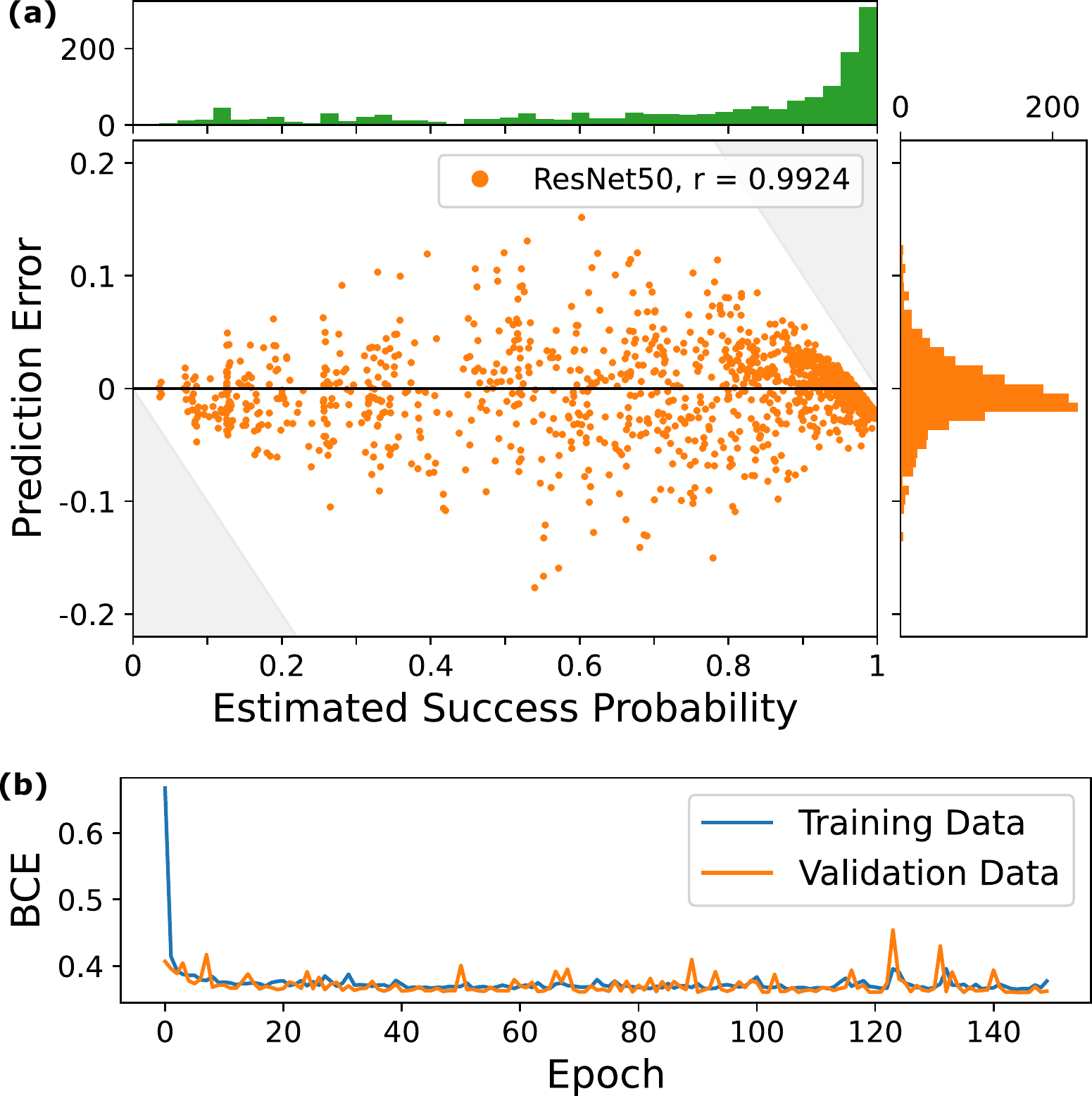}    \caption{\small{\textbf{Predicting circuit success probabilities using Resnet50.} We used transfer learning to create a capability model from ResNet50. \textbf{(a)} The prediction error on test data of our ResNet50-based capability model versus the true success probabilities (main plot), a histogram of the success probabilities (upper plot), and a histogram of prediction error (right plot). We observe moderate prediction accuracy on the test data ($\delta_{\textrm{abs}} = 2.48\%$). \textbf{(b)} Training and validation loss versus training epoch. The fluctuations in both the training and validation loss could be caused by a high learning rate ($\alpha = .0001$).}}
    \label{fig:resnet50-training-predictions}
\end{figure}

We trained our model for up to 150 epochs, using binary cross-entropy (BCE) on the training data as the loss metric. Throughout training, only the weights in the final single-unit dense layer were updated (all the weights in ResNet50 were frozen). Loss on the validation dataset was monitored, training was stopped after five epochs of increasing validation loss, and the final model used the weights that minimized the validation loss (this occurred after the 144\textsuperscript{th} epoch). Figure~\ref{fig:resnet50-training-predictions} (b) shows the training history.

The trained ResNet50-based capability model ($E_{\textrm{ResNet50}}$) has moderate prediction accuracy on the test data. Figure~\ref{fig:resnet50-training-predictions} (a) shows the prediction error of $E_{\textrm{ResNet50}}$ for every circuit $c$ in the test dataset, for which $\delta_{\textrm{abs}}(E_{\textrm{ResNet50}}) = 2.48\%$. This demonstrates the feasibility of using transfer learning to create capability models from large neural networks that have been trained for different prediction tasks. However, $E_{\textrm{ResNet50}}$'s prediction accuracy is worse than has been obtained using custom \emph{de novo} CNNs trained on data from this dataset (c.f., Ref.~\cite{Hot23} with $\delta_{\textrm{abs}} \approx 0.8\%$). This suggests that transfer learning is likely to prove most useful when neural networks that have been designed and trained as capability models need to make \emph{out-of-distribution} (OOD) predictions. Two particularly relevant examples of OOD prediction tasks are: (1) training on data from one circuit family (e.g., random circuits) and then making predicting for another circuit family (e.g., algorithm circuits), and (2) training on data from one processor (real or simulated) and then making predictions for a difference processor. Transfer learning has the potential to greatly improve OOD predictions by fine-tuning a pre-trained network using a small amount of new OOD training data.

\section{Conclusions}\label{sec:conclusions}
In this paper we have proposed a general framework for building predictive models from quantum computer benchmarking data. Our framework consists of fitting a \emph{capability model} to benchmarking data, and it can be applied to data from a wide range of benchmarks---including RB, cross-entropy benchmarking, algorithmic benchmarks, volumetric benchmarks, and the quantum volume benchmark. Capability models encompass a broad range of parameterized models, and we explored two interesting classes of capability model: ERMs (error rates models) and neural networks. 

ERMs are simple, flexible, interpretable, and scalable models that make accurate predictions when stochastic errors dominate. But even when a best-fit ERM's predictions have low accuracy, ERM's are still an powerful tool, because a best-fit ERM's parameters summarize the data in terms of effective error rates. In contrast, neural network capability models 
have the potential to be accurate in the presence of complex and poorly-understood errors~\cite{Hot23} but do not have interpretable parameters. In this work, we applied transfer learning to create a capability model from ResNet50, a large, pre-trained CNN. Although we obtained lower prediction accuracy than with bespoke models (see Ref.~\cite{Hot23}), our results demonstrate the promise and feasibility of a transfer learning approach to capability learning. 

\small{\section*{Data and code availability}
The data and code used in the project will be available at Ref.~\cite{dataset}. In the meantime, please email the corresponding author.

\section*{Acknowledgements}
This material was funded in part by the U.S. Department of Energy, Office of Science, Office of Advanced Scientific Computing Research, Quantum Testbed Pathfinder Program, and by the Laboratory Directed Research and Development program at Sandia National Laboratories. T.P. acknowledges support from an Office of Advanced Scientific Computing Research Early Career Award. Sandia National Laboratories is a multi-program laboratory managed and operated by National Technology and Engineering Solutions of Sandia, LLC., a wholly owned subsidiary of Honeywell International, Inc., for the U.S. Department of Energy's National Nuclear Security Administration under contract DE-NA-0003525. All statements of fact, opinion or conclusions contained herein are those of the authors and should not be construed as representing the official views or policies of the U.S. Department of Energy, or the U.S. Government. We acknowledge the use of IBM Quantum services for this work. The views expressed are those of the authors, and do not reflect the official policy or position of IBM or the IBM Quantum team.}

\bibliographystyle{IEEEtran}
\bibliography{IEEEabrv,main}

\end{document}